\documentclass[12pt,showpacs,preprintnumbers,amsmath,aps,amssymb]{revtex4}
\usepackage{graphicx}
\usepackage[english]{babel}
\usepackage{bm}
\usepackage[font=scriptsize]{caption}
\usepackage{ragged2e}
\usepackage[font=small,labelfont=bf,justification=raggedright]{caption}
\usepackage[utf8]{inputenc}
\usepackage[framemethod=tikz]{mdframed}

\begin{document}

\title{Non commutative classical and  Quantum fractionary Cosmology: FRW case. }

\author{J. Socorro$^{1}$ }
\email{socorro@fisica.ugto.mx}
\author{J. Juan Rosales $^2$}
\email{rosales@ugto.mx}
\author{Leonel Toledo-Sesma$^3$}
\email{ltoledos@ipn.mx} \affiliation{$^{1}$ Departamento de  F\'{\i}sica, DCI-Campus Le\'on, Universidad de Guanajuato,\\
C.P. 37150, Le\'on, Guanajuato, M\'exico\\
$^{2}$ Department of Electrical Engineering, Engineering Division
Campus Irapuato-Salamanca, University of Guanajuato,
Salamanca 36885, M\'exico.\\
$^{3}$ Instituto Polit\'ecnico Nacional, UPIIH, Carretera Pachuca -
Actopan Kil\'ometro 1+500, 42162. San Agust\'in Tlaxiaca, Hgo.
M\'exico}

\begin{abstract}
In this work we shall explore the effects of non commutativity in
fractional classical and quantum schemes using the flat
Friedmmann-Robertson-Walker (FRW) cosmological model coupled to a
scalar field in the K-essence formalism. In previous work we have
obtained the commutative solutions in both regimes into the
fractional framework. Here we introduce noncommutative variables,
considering that all minisuperspace variables $\rm q^i_{nc}$ do not
commute, so the symplectic structure was modified. In the quantum
regime, the probability density presents new structure in the scalar
field corresponding to the value of the non-commutative parameter,
in the sense that this probability density undergoes a shift back to
the direction of the scale factor, causing classical evolution to
arise earlier than in the commutative world.


\end{abstract}

  \maketitle                            
\section{introduction}
The study and applications of fractional calculus (FC) to cosmology
is a new line of research, that was born approximately twenty years
ago.  We have recently worked along this line in the theory of
K-essence due that in \cite{Socorro1} it is mentioned that by
quantifying different epochs of the K-essence theory, a fractional
Wheeler-DeWitt equation (WDW) in the scalar field component is
naturally obtained in different stages of the universe
\cite{universe,fra-fractionary}, however we have not found any work
in the literature, where the idea of non-commutativity (NC) is
applied to this formalism, which is why we are interested in
studying the effects of NC variables  from  the fractional calculus
approach, and seeing their effects on the exact solutions or
mathematical structure of the same. It is well known that there are
various ways to introduce non-commutativity in the phase space and
that they produce different dynamical systems from the same
Lagrangian \cite{Abreu-2006}, as can be shown for example in
reference \cite{De-andrade} and references cited therein.
 Therefore, distinct
choices for the NC algebra among the brackets render distinct
dynamic systems. We will use non-commutativity in the coordinate
space, which is where we have some working practice in the past,
leaving the application of moments space for the future,
\cite{Sabido,Aguero,Guzman,Ortiz,Socorro,Guzman-2011}, where other
quantities such as angular momentum appear between coordinates and
momenta \cite{sabido2018,sabido2024}.

Usually, K-essence models are restricted to the Lagrangian density
of the form~\cite{1,roland,chiba,bose,arroja,tejeiro}
\begin{equation}
 S=\int d^4x \, \sqrt{-g}\,\left[ f(\phi) \, {\cal G}(X)\right],
\end{equation}
where the canonical kinetic energy is given by ${\cal
G}(X)=X=-\frac{1}{2}\nabla_\mu \phi \nabla^\mu \phi$, $f(\phi)$ is
an arbitrary function of the scalar field $\phi$, and~g is the
determinant of the metric. K-essence was originally proposed as a
model for inflation, and then as~a model for dark energy, along with
explorations of unifying dark energy and dark matter. During the
development of research in non-commutative formalism within
fractional cosmology in k-essence theory, the presence of
non-commutativity that usually accompanied the term of the scale
factor, here is disrupted, since essentially Non-commutativity is
more present in the scalar field, modifying the mathematical
structure that usually occurs in works in this direction in other
non-fractional formalisms.

We construct the Lagrangian and Hamiltonian densities for the plane
FLRW cosmological model, considering a barotropic perfect fluid for
the scalar field in the variable $X$, and present the general case
in commutative (\ref{commutative}) and noncommutative formalism
(\ref{noncommutative}). We present the quantum version in both
cases,  in (\ref{quantum-commutative}) and
(\ref{quantum-noncommutative}), respectively. Finally, Section
(\ref{comment}) is devoted to~discussions.

\section{Commutative fractional classical exact
solution}\label{commutative}
 We start with the following classical
Lagrangian density that comes from the flat
Friedmann-Robertson-Walker fractionary cosmological model coupled to
a scalar field in the K-essence formalism \cite{universe}
\begin{equation}
 {\cal L} =  e^{3\Omega} \left[6 \frac{\dot \Omega^2 }{N}
  - \left(\frac{1}{2}\right)^{\alpha}\left( \dot \phi\right)^{2\alpha} N^{-2\alpha+1}
  \right].\label{lala}
\end{equation}

Using the standard definition  of the  momenta $
\Pi_{q^\mu}=\frac{\partial{\cal L}}{\partial{\dot q^\mu}}$, where $
q^{\mu}=(\Omega, \phi)$, we obtain
\begin{eqnarray}
 \Pi_\Omega&=& \frac{12}{N}e^{3\Omega}\dot \Omega, \quad \rightarrow \quad \dot \Omega=\frac{N}{12}e^{-3\Omega}\Pi_\Omega, \label{a} \\
 \Pi_\phi&=&
-\left(\frac{1}{2}\right)^\alpha\frac{2\alpha}{N^{2\alpha-1}}e^{3\Omega}{\dot
\phi}^{2\alpha -1}, \quad \rightarrow \quad \dot
\phi=-N\left[\frac{2^{\alpha-1}}{\alpha}e^{-3\Omega}\Pi_\phi\right]^{\frac{1}{2\alpha
-1}},\label{ph}
\end{eqnarray}
and introducing them into the Lagrangian density, we obtain the
canonical Lagrangian $ {\cal L}_{canonical}=\Pi_{q^\mu} \dot q^\mu
-N {\cal H}$ as
\begin{eqnarray}
 {\cal L}_{canonical}&=& \Pi_{q^\mu} \dot q^\mu
-\frac{N}{24}e^{-\frac{3}{2\alpha-1}\Omega} \left\{
e^{-\frac{6(\alpha-1)}{2\alpha-1}\Omega}\Pi_\Omega^2 -
\frac{12(2\alpha-1)}{\alpha}\,\Pi_\phi^{\frac{2\alpha}{2\alpha-1}}\right\}.
\label{canonical}
\end{eqnarray}
Performing the variation with respect to the lapse function $N$,
${\delta{\mathcal L}}_{canonical}/\delta N=0$, the~Hamiltonian
constraint $\mathcal H=0$ is obtained, where the classical density
is written as
\begin{equation}
 {\cal H}=\frac{1}{24}e^{-\frac{3}{2\alpha-1}\Omega} \left\{
e^{-\frac{6(\alpha-1)}{2\alpha-1}\Omega}\Pi_\Omega^2 -
\frac{12(2\alpha-1)}{\alpha}\left(\frac{2^{\alpha-1}}{\alpha}
\right)^{\frac{1}{2\alpha-1}}\,\Pi_\phi^{\frac{2\alpha}{2\alpha-1}}\right\}.
\label{hamia}
\end{equation}
For simplicity we work in the gauge
$N=24e^{\frac{3}{2\alpha-1}\Omega}$, and in the follow we use the
reduced Hamiltonian density,
\begin{equation}
{\cal H}= e^{-\frac{6(\alpha-1)}{2\alpha-1}\Omega}\Pi_\Omega^2 -
\frac{12(2\alpha-1)}{\alpha}\left(\frac{2^{\alpha-1}}{\alpha}
\right)^{\frac{1}{2\alpha-1}}\,\Pi_\phi^{\frac{2\alpha}{2\alpha-1}}.
\label{hami-alpha}
\end{equation}
In previous work \cite{universe}, we found that the barotropic
parameter in K-essence theory has the form
$\omega_X=\frac{1}{2\alpha-1}$, and the fractional parameter is
$\beta=\frac{2\alpha}{2\alpha -1}$, so, when $\omega_X \in [0,1]$,
thus $\beta \in [1,2]$, and when $\omega_X \in [-1,0)$, thus $\beta
\in (0,1]$. This is relevant
because in the quantum regime, 
the Laplace transform of a fractional
differential equation depends on the parameter $n=[\beta]$ (integer
part of the fractional parameter).

With this, the Hamiltonian density is rewritten as
\begin{equation}
{\cal H}= e^{-3(2-\beta)\Omega}\Pi_\Omega^2 -
\frac{24}{\beta}\left(\frac{2^{\alpha-1}}{\alpha}
\right)^{\frac{1}{2\alpha-1}}\,\Pi_\phi^{\beta}, \label{hami-beta}
\end{equation}
then, the Hamilton equations are
\begin{eqnarray}
\dot \Omega &=&2 e^{-3(2-\beta)\Omega}\Pi_\Omega, \label{dot-omega}\\
\dot \phi &=& -24 \left(\frac{2^{\alpha-1}}{\alpha}
\right)^{\frac{1}{2\alpha-1}}\,\Pi_\phi^{\beta-1}, \label{dot-phi}\\
\dot \Pi_\Omega &=& 3(2-\beta) e^{-3(2-\beta)\Omega}\Pi_\Omega^2,
\label{dot-pi-omega}\\
\dot \Pi_\phi &=& 0, \qquad \Pi_\phi=p_\phi=constant.
\label{dot-pi-phi}
\end{eqnarray}
substituting these results in the Hamiltonian constraint, we have
that
\begin{equation}
\dot \Pi_\Omega = p_\beta, \qquad \Pi_\Omega(t)=p_0+p_\beta (t-t_0),
 \label{sol-pi-omega}
\end{equation}
where $p_0$ is an integration constant and
$p_\beta=\frac{72(2-\beta)}{\beta}\left(\frac{2^{\alpha-1}}{\alpha}
\right)^{\frac{1}{2\alpha-1}}\,p_\phi^{\beta}$. With this  and using
the equation (\ref{dot-omega}), the solution for the scale factor
$A(t)=e^{\Omega}$ becomes,
\begin{equation}
A(t)=\left[A_0 + 6(2-\beta)p_0(t-t_0) + 3(2-\beta)p_\beta (t-t_0)^2
\right]^{\frac{1}{3(2-\beta)}}, \label{scale-factor}
\end{equation}
and the solution for the scalar field $\phi$ is
\begin{equation}
\phi(t)=\phi_0-24 \left(\frac{2^{\alpha-1}}{\alpha}
\right)^{\frac{1}{2\alpha-1}}\,p_\phi^{\beta-1} (t-t_0).
\label{solution-phi}
\end{equation}

\section{Noncommutative fractional classical exact solution}\label{noncommutative}
 We start with the following classical
hamiltonian that comes from the flat Friedmann-Robertson-Walker
fractionary cosmological model coupled to a scalar field in the
K-essence formalism (\ref{hami-beta}), written in term of the
fractional parameter $\beta=\frac{2\alpha}{2\alpha-1}$ and in
particular gauge, where in order to find the commutative equation of
motion, we use the classical phase space variables $\rm
q^\mu=(\Omega,\phi)$, where the Poisson algebra for these
minisuperspace variables are
\begin{equation}
 \left\{ q^\mu, q^\nu  \right\}=0 \qquad \left\{ \Pi_{q^\mu}, \Pi_{q^\nu} \right\}=0, \qquad \left\{q^\mu,\Pi_{q^\nu}
\right\}=\delta^\mu_\nu, \label{cbracket}
\end{equation}

\begin{equation}
 {\cal H}=\frac{1}{24}e^{3(1-\beta)\Omega} \left\{
e^{-3(2-\beta)\Omega}\Pi_\Omega^2 -
\frac{24}{\beta}\left(\frac{2^{\alpha-1}}{\alpha}
\right)^{\frac{1}{2\alpha-1}}\,\Pi_\phi^{\beta}\right\}.
\label{hami-n}
\end{equation}
In the commutative model the solutions to the Hamiltonian equations
are the same as in General Relativity, modified only by the
fractional parameter. Now the natural extension is to consider the
noncommutative version of our model, with the idea of
non-commutative between the two variables $\rm
(\Omega_{nc},\phi_{nc})$, so we apply a deformation of the Poisson
algebra. For this, we start with the usual Hamiltonian
(\ref{hami-beta}), but the symplectic structure is modify as follow
\begin{equation}
\left\{\Pi_{\Omega},\Pi_\phi\right\}_\star =0, \qquad
\left\{q^\mu,\Pi_{q^\mu} \right\}_\star = 1, \qquad
\left\{\Omega,\phi\right\}_\star = \theta, \label{ncbracket}
\end{equation}
where the $\star$ is the Moyal product \cite{Szabo2}, and  the
resulting Hamiltonian density is
\begin{equation}
 {\cal H}_{nc}=
e^{-3(2-\beta)\Omega_{nc}}\Pi_\Omega^2 -
\frac{24}{\beta}\left(\frac{2^{\alpha-1}}{\alpha}
\right)^{\frac{1}{2\alpha-1}}\,\Pi_\phi^{\beta}, \label{hami-nc}
\end{equation}
but the symplectic structure is the one that we know, the
commutative one (\ref{cbracket}). It is well known that, there are
two formalisms to study the non-commutative equations of motion, for
the first formalism that we exposed has the original variables, but
with the modified symplectic structure,
\begin{eqnarray}
\dot{q^\mu_{nc}}&=&\rm \{q^\mu, {\cal H}\}_\star ,
\nonumber\\
\dot{\Pi^\mu_{nc}}&=&\rm \{P^\mu, {\cal H}\}_\star ,
\end{eqnarray}
and for the second formalism we use the shifted variables (Bopp
shift approach) but with the original (commutative) symplectic
structure
\begin{eqnarray}
\dot{q^\mu_{nc}}&=&\rm \{q^\mu_{nc}, {\cal H}_{nc}\} ,
\nonumber\\
\dot{\Pi^\mu_{nc}}&=&\rm \{\Pi^\mu_{nc}, {\cal H}_{nc}\} ,
\end{eqnarray}
in  both approaches, we have the same result.

The commutation relations (\ref{cbracket}) can be implemented in
terms of the commuting coordinates of the standard quantum mechanics ( Bopp shift) and it results in a modification of the potential like
term of the Hamiltonian density and one possibility  is, for
example,
\begin{equation}
\rm \Omega_{nc} = \Omega + \frac{\theta}{2}\Pi_\phi, \qquad
\phi_{nc} = \phi -\frac{\theta}{2}\Pi_\Omega
 \label{transformation}
\end{equation}
These transformations are not the most general possible to define
non-commutative fields. With this in mind, our hamiltonian density
has the form
\begin{equation}
 {\cal H}_{nc}=
e^{-3(2-\beta)\left[\Omega +\frac{\theta}{2}
\Pi_\phi\right]}\Pi_\Omega^2 -
\frac{24}{\beta}\left(\frac{2^{\alpha-1}}{\alpha}
\right)^{\frac{1}{2\alpha-1}}\,\Pi_\phi^{\beta}, \label{hami-nc-n}
\end{equation}
the Hamilton equations are
\begin{eqnarray}
\dot \Omega &=& 2e^{-3(2-\beta)[\Omega+\frac{\theta}{2}\Pi_\phi]}\Pi_\Omega, \label{omega}\\
\dot \phi &=&
-\frac{3\theta(2-\beta)}{2}e^{-3(2-\beta)[\Omega+\frac{\theta}{2}\Pi_\phi]}\Pi_\Omega^2-24\left(\frac{2^{\alpha-1}}{\alpha}
\right)^{\frac{1}{2\alpha-1}}\,\Pi_\phi^{\beta-1}, \label{phi}\\
\dot \Pi_\phi &=& 0, \qquad \dot \Pi_\Omega
=3(2-\beta)e^{-3(2-\beta)[\Omega+\frac{\theta}{2}\Pi_\phi]}
\Pi^2_\Omega. \label{momenta}
\end{eqnarray}
with these equations, the solution for $\Pi_\Omega$ is the same as
in the commutative case, so the solution for the scale factor
becomes
\begin{equation}
A_\theta(t)=e^{-\frac{\theta}{2}p_\phi} A(t),
\end{equation}
where $A(t)$ is the solution presented in equation
(\ref{scale-factor}). The solution for the scalar field is related
with the $\Pi_\Omega$ as
$$\dot \phi=-\frac{\theta}{2}\dot \Pi_\Omega-24 \left(\frac{2^{\alpha-1}}{\alpha}
\right)^{\frac{1}{2\alpha-1}}\,p_\phi^{\beta-1},$$ then
\begin{equation}
\phi(t)=\phi_0-24 \left(\frac{2^{\alpha-1}}{\alpha}
\right)^{\frac{1}{2\alpha-1}}\,p_\phi^{\beta-1}(t-t_0)
-\frac{\theta}{2}\Pi_\Omega,
\end{equation}
for both commutative solutions, the scale factor y scalar field
$\phi$ are obtained when the non-commutative parameter $\theta$,
goes to zero.

\section{Commutative fractional quantum exact solution}\label{quantum-commutative}
 The Wheeler-DeWitt (WDW)
equation for this model is obtained  by making the usual
substitution $ \Pi_{q^\mu}=-i \hbar \frac{\partial}{\partial q^\mu}$
into (\ref {hami-beta}) and promoting the classical Hamiltonian
density in the differential operator, applied to the wave function
$\Psi(\Omega,\phi)$, $\hat{\cal H}\Psi=0$. Then, we have
\begin{equation}
-\hbar^2 e^{-3(2-\beta)\Omega}\frac{\partial^2 \Psi}{\partial
\Omega^2}-\frac{24}{\beta}\,\hbar^{\beta}
\left(\frac{2^{\alpha-1}}{\alpha}
\right)^{\frac{1}{2\alpha-1}}\frac{\partial^\beta \Psi}{\partial
\phi^\beta}\,=0. \label{q-wdw}
\end{equation}
For simplicity, the  factor $ e^{-3(2-\beta)\Omega}$ may be the
factor ordered with $ \hat \Pi_\Omega$ in many ways. Hartle and
Hawking~\cite{HH} suggested what  might be called semi-general
factor ordering, which, in~this case, would order the terms  $
e^{-3(2-\beta)\Omega} \hat \Pi^2_\Omega$ as $ - \hbar^2 e^{-[
3(2-\beta)- Q]\Omega}\,
\partial_\Omega e^{-Q\Omega}
\partial_\Omega= -\hbar^2\, e^{-3(2-\beta)\Omega}\, \partial^2_\Omega + \hbar^2\, Q\,
e^{-3(2-\beta)\Omega} \partial_\Omega$, where $Q$ is any real
constant that measures the ambiguity in the factor ordering in the
variables $ \Omega$  and its corresponding momenta.

Thus, the equation (\ref{q-wdw}) is rewritten as
\begin{equation}
\hbar^2 e^{-3(2-\beta)\Omega}\left[-\frac{\partial^2 \Psi}{\partial
\Omega^2} + Q \frac{\partial \Psi}{\partial
\Omega}\right]-\frac{24}{\beta}\,\hbar^{\beta}
\left(\frac{2^{\alpha-1}}{\alpha}
\right)^{\frac{1}{2\alpha-1}}\frac{\partial^\beta \Psi}{\partial
\phi^\beta}\,=0. \label{wdw-factor-ordering-a}
\end{equation}
By employing the separation variables method for the wave function
$\Psi={\cal A}(\Omega)\, {\cal B}(\phi)$, we have the following two
differential equations for $(\Omega,\phi)$
\begin{eqnarray}
 \frac{d^2 {\cal A}}{d\Omega^2}\, - Q \frac{d {\cal A}}{d \Omega}
\mp \frac{\mu^2}{\hbar^2}e^{3(2-\beta)\Omega}{\cal A}
 &=& 0,\label{scale}\\
 \frac{d^\beta {\cal B_\pm}}{ d \phi^\beta} \pm
\left(\frac{\alpha}{2^{\alpha-1}} \right)^{\frac{1}{2\alpha-1}}
\frac{\mu^2\,\beta}{24 \hbar^\beta} {\cal B}_\pm&=&0,\label {phi-1}
\end{eqnarray}
where $ {\cal B}_\pm$ considers the sign in the differential
equation. The~fractional differential Equation~(\ref{phi-1}) can be
given in the fractional frameworks,
following~\cite{Rosales1,Rosales2} and identifying $
\gamma=\frac{\beta}{2}=\frac{\alpha}{2\alpha-1},$ where now, $
\gamma$ is the order of the fractional derivative taking values in $
0<\gamma\leq 1$; then, we can write
\begin{equation}
  \frac{d^{2\gamma} {\cal B}_\pm}{d \phi^{2\gamma}} \pm
\left(\frac{\alpha}{2^{\alpha-1}}
\right)^{\frac{1}{2\alpha-1}}\frac{\gamma \mu^2 }{12
\hbar^{2\gamma}} {\cal B}_\pm=0 ,\qquad 0<\gamma \leq 1,
\label{phi433}
\end{equation}
the solution of the Equation~(\ref{phi433}) with a positive sign may
be obtained by applying direct and inverse Laplace
transforms~\cite{Rosales2}, providing
\begin{equation}
 {\cal B}_+ (\phi, \gamma) =
 \mathbb{E}_{2\gamma}\left(- z^2\right),
 \qquad z=\left(\frac{\alpha}{2^{\alpha-1}}
\right)^{\frac{1}{2(2\alpha-1)}}\frac{\sqrt{\gamma} \mu
}{2\sqrt{3}\hbar^{\gamma}} \phi^\gamma,\qquad 0<\gamma \leq 1.
\label{R}
 \end{equation}

In the ordinary case, $\gamma=1$; then, the~solution
is~\cite{Rosales2},
\begin{equation}
{\cal B}_+ (\phi, 1) =\mathbb{E}_2\left[-
\left(\frac{\mu}{2\sqrt{3}\hbar} (\phi-\phi_0) \right)^2\right]=
cos\left(\frac{\mu}{2\sqrt{3}\hbar} (\phi-\phi_0) \right),
\label{clasico1}
\end{equation}

 Following the book of Polyanin~\cite{polyanin} (page 179.10), we discovered
the solution for the first equation for $\gamma\not=1$, considering
different values in the factor ordering parameter and both signs in
the constant $\mu^2$,
\begin{equation}  {\cal A}= A_0\, e^{\frac{
Q\Omega}{2}}\,Z_\nu\left[\frac{2\sqrt{\pm \mu^2}}{3\hbar(2-\beta)}
e^{\frac{3(2-\beta)}{2}\Omega} \right]= \left\{ \begin{tabular}{ll}
$A_0\, e^{\frac{
Q\Omega}{2}}\,K_\nu\left[\frac{\mu}{3\hbar(1-\gamma)}
e^{3(1-\gamma)\Omega} \right],$ & \quad to\,\,$-\mu^2$,\\
$A_0\, e^{\frac{
Q\Omega}{2}}\,J_\nu\left[\frac{\mu}{3\hbar(1-\gamma)}
e^{3(1-\gamma)\Omega} \right],$ & \quad to\,\, $+\mu^2$
\end{tabular} \right. \label{a-solution}
\end{equation}
with order $  \nu=\pm \frac{Q}{6(1-\gamma)}$, where we had written
the second expression in terms of the fractional order
$\gamma=\frac{\beta}{2}$, and~the solutions which become dependent
on the sign of its argument; when $\sqrt{\mu^2}$ (for ${\cal B}_-$),
the~Bessel function $Z_\nu$ becomes the ordinary Bessel function
$J_\nu$. When $\sqrt{-1\mu^2}$ (for ${\cal B}_+$), this becomes the
modified Bessel function $K_\nu$. For~the particular values
$\beta=2$ ($\gamma=1$), it will be necessary to solve the original
differential equation for this~variable.

Then, we have the probability density  $ |\Psi|^2$ by considering
only $ {\cal B}_+$, $\gamma \not= 1$,
\begin{equation}
 |\Psi|^2=\psi_0^2 \,e^{Q\Omega}\, \mathbb{E}^2_{2\gamma}\left(-
z^2\right)\,\,K_\nu\left[\frac{\mu}{3\hbar(1-\gamma)}
e^{3(1-\gamma)\Omega} \right], \quad
z=\left(\frac{\alpha}{2^{\alpha-1}}
\right)^{\frac{1}{2(2\alpha-1)}}\frac{\sqrt{\gamma} \mu
}{2\sqrt{3}\hbar^{\gamma}} \phi^\gamma. \label{dens}
\end{equation}

We will now report the solution for the $\beta=2, \to \gamma=1$
case, which we have not reported before, considering the minus sign
in the constant $\mu^2$, the general solution for the function
${\cal A}$ becomes
\begin{equation}
{\cal A}= e^{\frac{Q}{2}\Omega}\left[A_0e^{r_1 \Omega}+ B_0
e^{-r_1\Omega} \right], \qquad
r_1=\frac{1}{2}\sqrt{Q^2+4\frac{\mu^2}{\hbar^2}},
\end{equation}
and for the other sign $+\mu^2$, becomes
\begin{equation}
{\cal A}= e^{\frac{Q}{2}\Omega}\left[A_0e^{r_2 \Omega}+ B_0
e^{-r_2\Omega} \right], \qquad
r_2=\frac{1}{2}\sqrt{Q^2-4\frac{\mu^2}{\hbar^2}},
\end{equation}
and the corresponding solutions to (\ref{phi433}) for both signs are
\begin{equation}
{\cal B}_\pm=\left\{ \begin{tabular}{l} $Cos\left(
\frac{\mu}{2\hbar\sqrt{3
}}(\phi-\phi_0)\right)$\\
$Cosh\left( \frac{\mu}{2\hbar\sqrt{3 }}(\phi +\phi_0) \right)$
\end{tabular} \right.
\end{equation}
so, the probability density becomes
\begin{equation}
|\Psi|^2=\left\{ \begin{tabular}{ll} $Cos^2\left(
\frac{\mu}{2\hbar\sqrt{3
}}(\phi-\phi_0)\right)e^{Q\Omega}\left[A_0e^{r_1 \Omega}+
B_0 \,e^{-r_1\Omega} \right]^2,$&\quad for $-\mu^2$\\
$Cosh^2\left( \frac{\mu}{2\hbar\sqrt{3 }}(\phi +\phi_0)
\right)e^{Q\Omega}\left[A_0e^{r_2 \Omega}+ B_0 e^{-r_2\Omega}
\right]^2,$& \quad for $+\mu^2$
\end{tabular} \right.
\end{equation}

 On the other hand, it is well-known that in the standard
quantum cosmology, the wave function is unnormalized. There is no
systematic method to do this, as the Hamiltonian density is not
Hermitian. In~particular cases, wave packets can be constructed, and
from these wave packets we can construct a normalized wave function.
In~this work, we could not construct these wave packets. We hope to
be able to do it in future~studies.

\section{Non-commutative fractional quantum exact solution} \label{quantum-noncommutative}
As already mentioned, we are looking for the non-commutative
deformation of the flat FRW quantum cosmological model. In order to
find the non-commutative generalization, we need to solve the
non-commutative Einstein equation, this is a formidable task due to
the highly non linear character of the theory, fortunately we can
circumvent these difficulties by following Ref. \cite{ncqc}.

Now we can proceed to the non-commutative model, we will consider,
that the minisuperspace  variables $\rm q^i=(\Omega,\phi)$ do not
commute, so that the symplectic structure is modified as follows
\begin{equation}
\rm [q^i, q^j ]= i\theta^{ij}, \quad [\hat \Pi_i, \hat\Pi_j]=0,
\quad [q^i,\hat \Pi_j]=i\delta^i_j, \label{rules-c}
\end{equation}
in particular, we choose the following representation
\begin{eqnarray}
\rm [\Omega,\phi]&=&\rm i\theta, \label{rules1}
\end{eqnarray}
where the $\rm \theta$ parameters are a measure of the non
commutativity between the minisuperspace variables. The commutation
relations (\ref{rules-c}) or (\ref{rules1}) are not the most general
ones to define a non-commutative field.

We consider the non-commutative hamiltonian density in a simple way,
as
\begin{equation}
\hbar^2 e^{-3(2-\beta)\Omega_{nc}}\left[-\frac{\partial^2
\Psi}{\partial \Omega^2} + Q \frac{\partial \Psi}{\partial
\Omega}\right]-\frac{24}{\beta}\,\hbar^{\beta}
\left(\frac{2^{\alpha-1}}{\alpha}
\right)^{\frac{1}{2\alpha-1}}\frac{\partial^\beta \Psi}{\partial
\phi^\beta}\,=0. \label{wdw-factor-ordering}
\end{equation}

It is well known that this non-commutativity can be formulated in
term of non-commutative minisuperspace functions with the Moyal star
product $\star$ of functions. The commutation relations
(\ref{rules-c}) can be implemented in terms of the commuting
 coordinates of the standard quantum mechanics ( Bopp shift) and it results in a
 modification of the potential like term of
the WDW equation \cite{ncqc,pimentel}, and one possibility  is, for
example,
\begin{equation}
 \Omega_{nc} \to  \Omega + \frac{\theta}{2}\hat\Pi_\phi, \qquad
 \phi_{nc} \to \phi-\frac{\theta}{2}\hat\Pi_\Omega
 \label{transformation-q}
\end{equation}
These transformation are not the most general possible to define
noncommutative fields,

However, these shifts  modify the potential term in the following
way
\begin{equation}
\hbar^2 e^{-3(2-\beta)[\Omega+\frac{\theta}{2}\hat
\Pi_\phi]}\left[-\frac{\partial^2 \Psi}{\partial \Omega^2} + Q
\frac{\partial \Psi}{\partial
\Omega}\right]-\frac{24}{\beta}\,\hbar^{\beta}
\left(\frac{2^{\alpha-1}}{\alpha}
\right)^{\frac{1}{2\alpha-1}}\frac{\partial^\beta \Psi}{\partial
\phi^\beta}\,=0. \label{wdw-nc}
\end{equation}
As in the commutative case, we choose the wave function to be
separable, $\Psi(\Omega,\phi)={\cal A}(\Omega)\, {\cal C}(\phi)$,
getting
\begin{equation}
e^{-i\hbar \frac{\theta}{2}\frac{d}{d\phi}} {\cal C}\left[\hbar^2
e^{-3(2-\beta)\Omega}\left( -\frac{d^2 {\cal A}}{d \Omega^2}+ Q\,
\frac{d {\cal A}}{d \Omega} \right) \right] - {\cal A}
\frac{24}{\beta}\,\hbar^{\beta} \left(\frac{2^{\alpha-1}}{\alpha}
\right)^{\frac{1}{2\alpha-1}}\frac{d^\beta {\cal C}}{d \phi^\beta}
\end{equation}
which can be rewritten as
\begin{equation}
e^{-i\hbar \frac{\theta}{2}\frac{d}{d\phi}} {\cal C}\left[\hbar^2
e^{-3(2-\beta)\Omega}\frac{\left( -\frac{d^2 {\cal A}}{d \Omega^2}+
Q\, \frac{d {\cal A}}{d \Omega} \right)}{{\cal A}} \right] -
\frac{24}{\beta}\,\hbar^{\beta} \left(\frac{2^{\alpha-1}}{\alpha}
\right)^{\frac{1}{2\alpha-1}}\frac{d^\beta {\cal C}}{d \phi^\beta},
\label{master}
\end{equation}
if we want this equation to be separable, we must choose to make the
term within the square parentheses [\,] a constant, in particular
$\mp \mu^2$, with this choice, we retrieve the commutative quantum
equation  for the function ${\cal A}$, (\ref{scale}), with the same
quantum solution (\ref{a-solution}).

At this point we want to note that in commutative quantum cosmology,
the prefactor that accompanies its moments is not contemplated when
we use a particular gauge, and usually the non-commutative parameter
enters the solution of the $\Omega$ function, not that of scalar
field $\phi$. In this case, the appearance of the prefactor in
fractional cosmology makes the solution in $\Omega$ remain the same,
but not the part of the scalar field, where the non-commutative
parameter appears and the mathematical structure is completely
different.

That said, the expression (\ref{master}) becomes
\begin{equation}
\mp \mu^2 e^{-i\hbar \frac{\theta}{2}\frac{d}{d\phi}} {\cal C} -
\frac{24}{\beta}\,\hbar^{\beta} \left(\frac{2^{\alpha-1}}{\alpha}
\right)^{\frac{1}{2\alpha-1}}\frac{d^\beta {\cal C}}{d \phi^\beta},
\label{master-n}
\end{equation}
since the non-commutative parameter $\theta$ is very small, we can
stay until the first term in this one, obtaining
\begin{equation}
\frac{d^\beta {\cal C_\pm}}{ d \phi^\beta} \pm
\left(\frac{\alpha}{2^{\alpha-1}} \right)^{\frac{1}{2\alpha-1}}
\frac{\mu^2\,\beta}{24 \hbar^\beta} {\cal C}_\pm \mp \hbar \theta
\left(\frac{\alpha}{2^{\alpha-1}} \right)^{\frac{1}{2\alpha-1}}
\frac{i\mu^2\,\beta}{48 \hbar^\beta} \frac{d{\cal
C}_\pm}{d\phi}=0,\label {phi-n}
\end{equation}
in this fractional differential equation, when $\theta=0$ we recover
the commutative equation for the quantum function ${\cal C} = {\cal
B}$ (\ref{phi-1}). Now we solve the equation (\ref{phi-n}) written
as follows

\begin{equation}
\frac{d^\beta {\cal C_\pm}}{ d \phi^\beta} \mp \frac{i\, \hbar
\theta}{2} q_{\alpha,\beta}\frac{d{\cal C}_\pm}{d\phi} \pm
q_{\alpha,\beta} {\cal C}_\pm =0, \label{ma-ster}
\end{equation}
where $q_{\alpha,\beta}= \left(\frac{\alpha}{2^{\alpha-1}}
\right)^{\frac{1}{2\alpha-1}} \frac{\mu^2\,\beta}{24 \hbar^\beta}$.
For the particular value $\beta=2$, we can observe that the equation
(\ref{wdw-factor-ordering}), becomes the ordinary commutative
quantum equation, then the quantum solutions, commutative and
non-commutative, are the same in this approach to k-essence theory.

However, in the dust scenario $(\beta = 1, \alpha\to \infty),
q_{\infty,1}=\frac{\sqrt{2}\mu^2}{48\hbar}$. Equation
\eqref{ma-ster} takes the form

\begin{equation}
z_{0}\frac{d\mathcal{C}_{\pm}}{d\phi} \pm \mathcal{C}_{\pm} = 0,
\qquad  z_{0} = \frac{48\hbar}{\sqrt{2}\mu^{2}} \mp
i\frac{\hbar\theta}{2}, \label{dust1}
\end{equation}
whose solution is given by

\begin{equation}
\mathcal{C}_{\pm} = \eta_{\pm}\, e^{\mp\,\varepsilon\Delta\phi}\,
e^{- i\,\theta\psi\Delta\phi}, \qquad \varepsilon =
\frac{96\sqrt{2}\hbar\mu^{2}}{\mu^{4}\hbar^2\theta^{2} +
4608\hbar^{2}}, \quad \psi = \frac{2\hbar
\mu^{4}}{\mu^{4}\hbar^2\theta^{2} + 4608\hbar^{2}}. \label{dusfinal}
\end{equation}
Thus, the probability density becomes (considering only the real
part of the complex exponential in $\theta$)

\begin{equation}  \Psi^2 = \left\{ \begin{tabular}{ll}
$A_0\, e^{-2\varepsilon\Delta\phi}\,
Cos^2\left[\theta\psi\Delta\phi\right]\, e^{
Q\Omega}\,K_\nu^2\left[\frac{2\mu}{3\hbar}
e^{\frac{3}{2}\Omega} \right],$ & \quad to\,\,$-\mu^2$,\\
$A_0\, e^{2\varepsilon\Delta\phi}\,
Cos^2\left[\theta\psi\Delta\phi\right]\,e^{ Q\Omega}\,J_\nu^2
\left[\frac{2\mu}{3\hbar} e^{\frac{3}{2}\Omega} \right],$ & \quad
to\,\, $+\mu^2$
\end{tabular} \right. \label{solution-both}
\end{equation}

\begin{figure}[ht]
\includegraphics[scale=0.6]{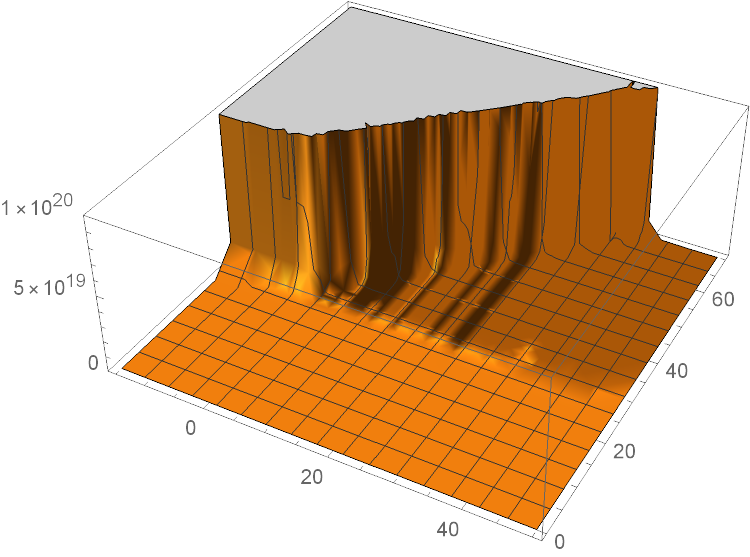}
\includegraphics[scale=0.6]{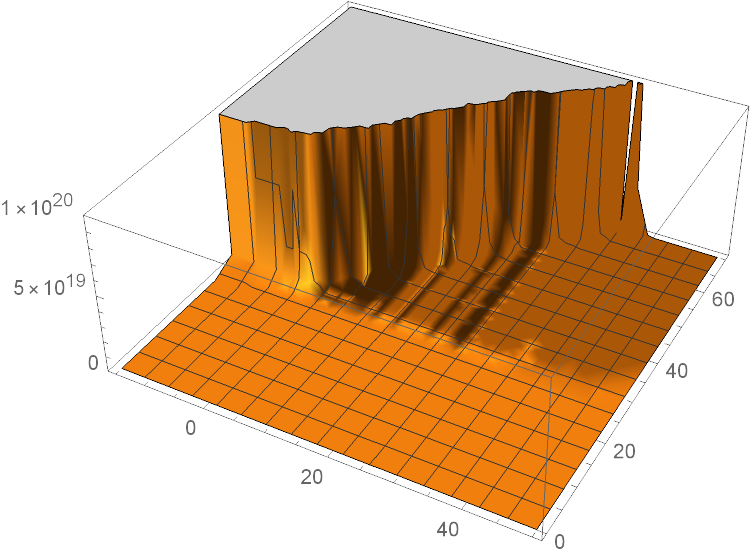}
\includegraphics[scale=0.6]{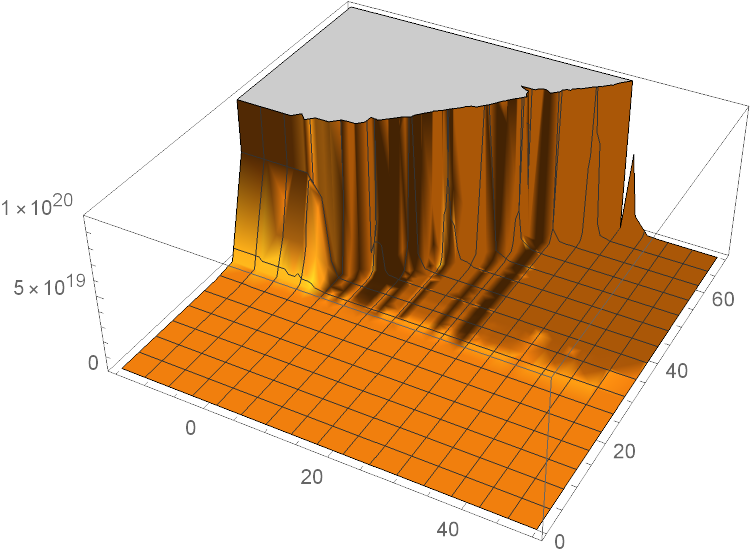}
\includegraphics[scale=0.6]{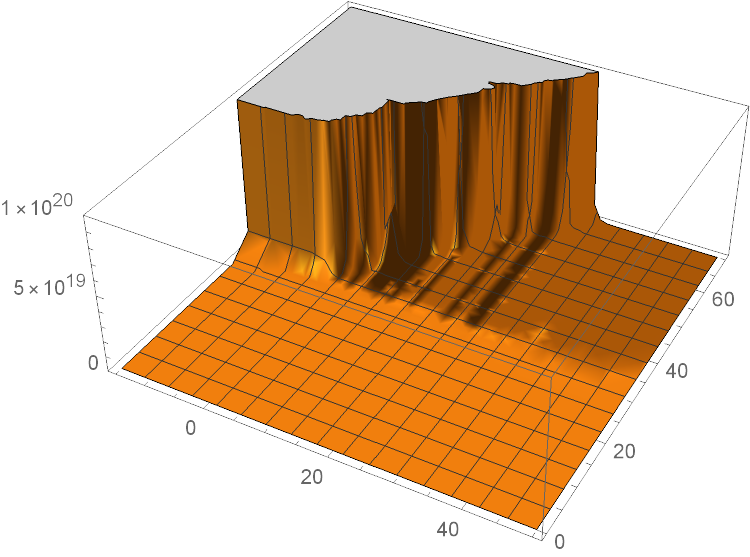}

\caption{In the following plots, we show the behavior of the
probability density of equation (\ref{solution-both}), considering
the sign in $+\mu^2$, taking the values $\mu=5$, $Q=0$ and
$\theta=0,0.1, 0.5, 1$, respectively.}\label{c-}
\end{figure}

To make the figure \ref{c-}, we use the ordinary Bessel function. We
can see the effect of the combination of the parameters $\theta$ and
$\mu$, where the probability density undergoes a shift in the
behavior of the scalar field, at the beginning and at the end, that
is, modifying the structure. As we can see, at $\theta=0$, a crack
appears, at $\theta=0.1$, it separates and a peak appears, at
$\theta=0.5$, the peak decreases and disappears at $\theta=1$, when
$\mu=$5. However, the fact that some peaks no longer appear does not
mean that they have been cancelled,
 but rather that, due to the change in probability density, the scales of these peaks are no longer on the graph.


\begin{figure}[ht]
\includegraphics[scale=0.6]{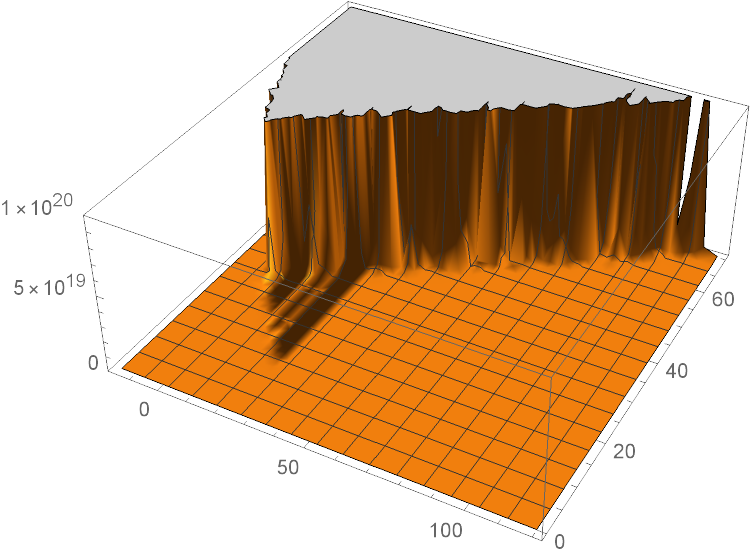}
\includegraphics[scale=0.6]{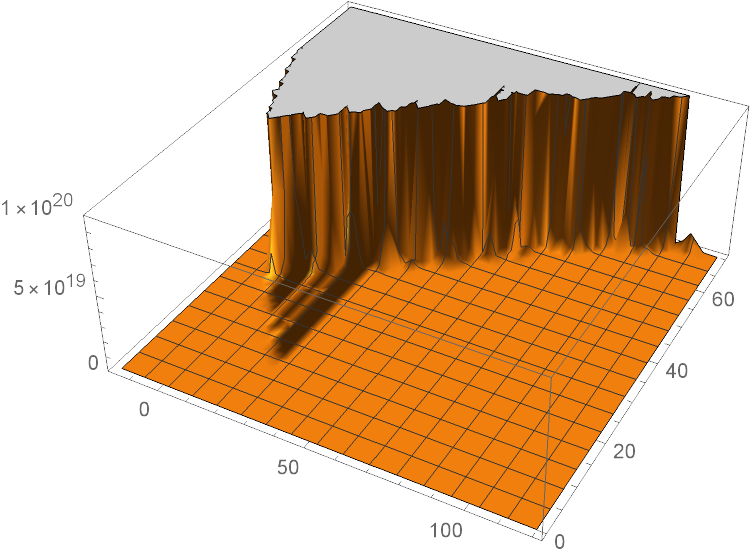}
\includegraphics[scale=0.6]{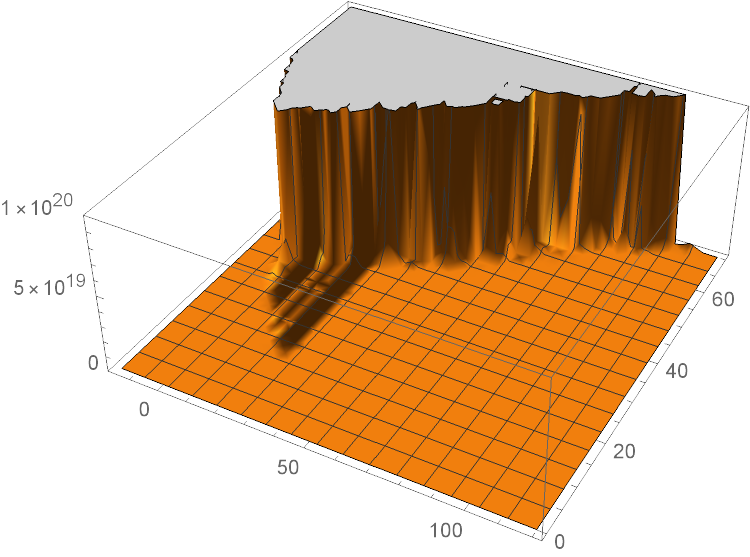}
\includegraphics[scale=0.6]{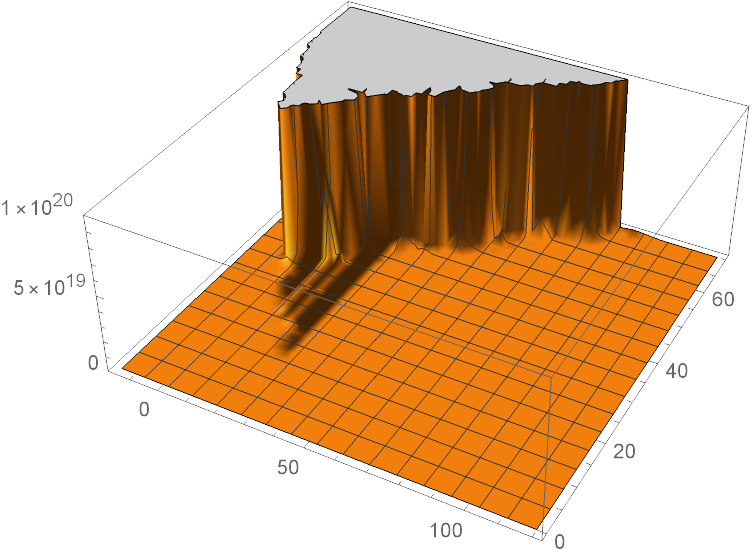}

\caption{In the following plots we show the behavior of the
probability density of equation (\ref{solution-both}), considering
the sign in $+\mu^2$, taking the values $\mu=5$, $Q=1$ and
$\theta=0,0.1, 0.5, 1$, respectively.}\label{c-1}
\end{figure}

\begin{figure}[ht]
\includegraphics[scale=0.6]{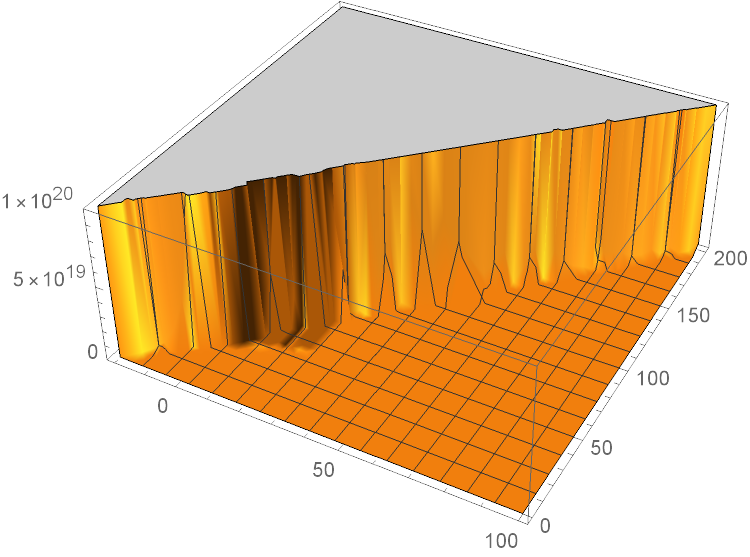}
\includegraphics[scale=0.6]{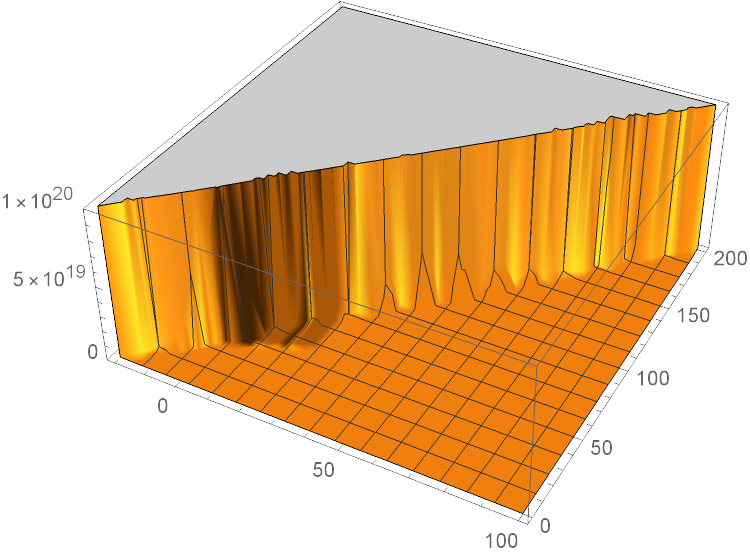}
\includegraphics[scale=0.6]{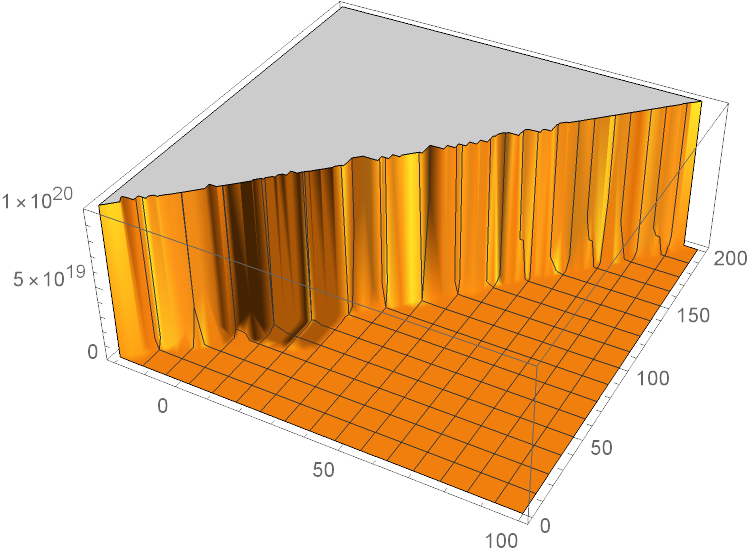}
\includegraphics[scale=0.6]{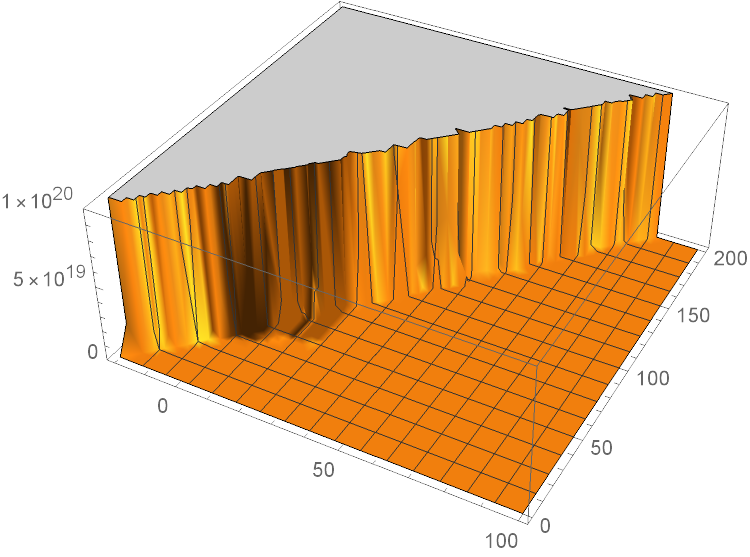}

\caption{In the following plots we show the behavior of the
probability density of equation (\ref{solution-both}), considering
the sign in $+\mu^2$, taking the values $\mu=5$, $Q=-1$ and
$\theta=0,0.1, 0.5, 1$, respectively.}\label{c-11}
\end{figure}

In the figure \ref{c-1}, the previously mentioned pattern is
repeated when the factor ordering parameter is $Q=1$, but more
noticeably in the shift towards the origin of the $\Omega$ variable.
In the figure \ref{c-11} for the factor ordering $Q=-1$, the shift
is slower, but persists.

\begin{figure}[ht]
\includegraphics[scale=0.6]{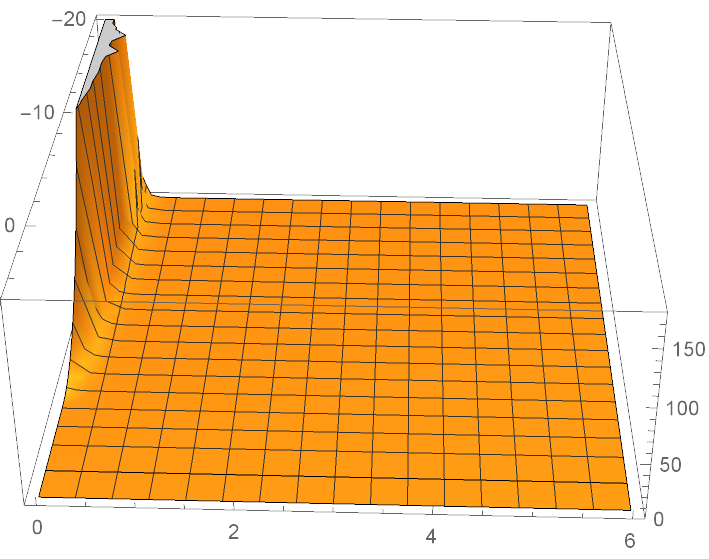}
\includegraphics[scale=0.6]{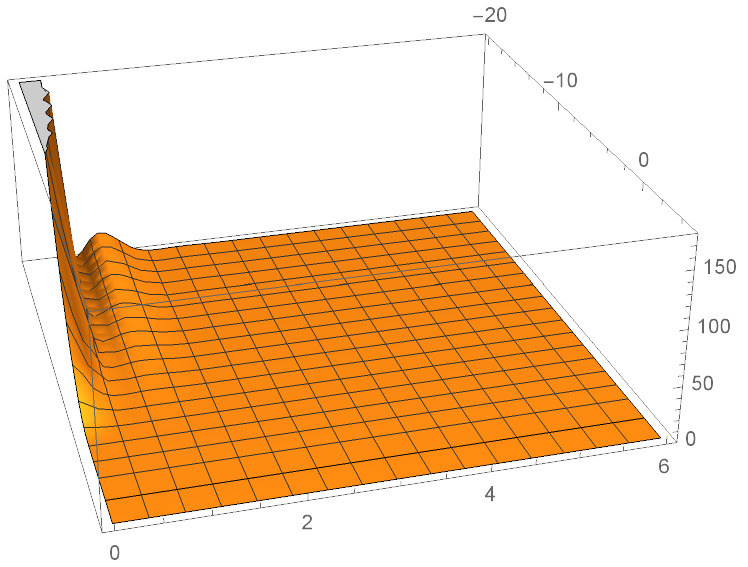}
\includegraphics[scale=0.6]{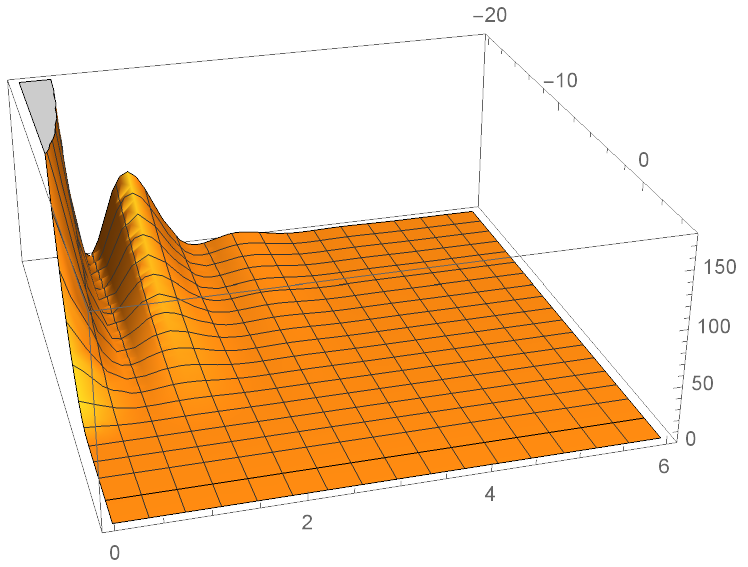}
\includegraphics[scale=0.6]{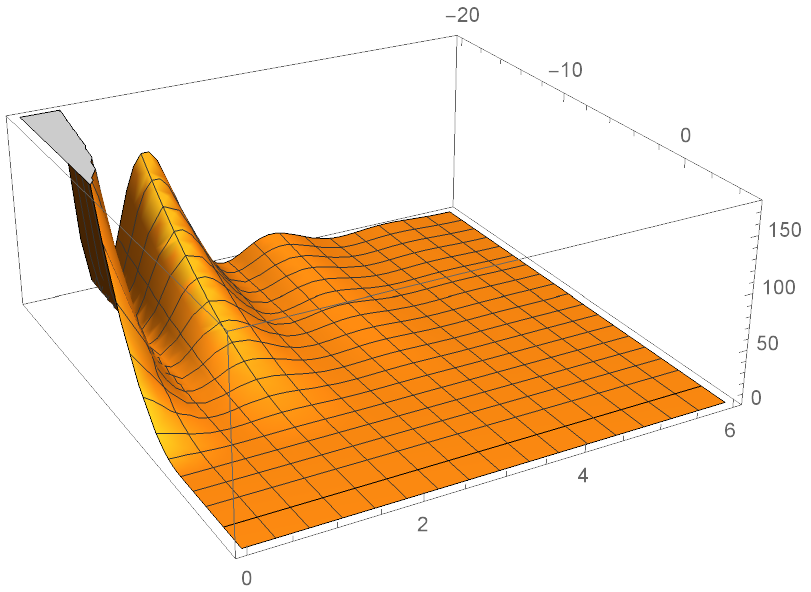}
\caption{In the following plots, we show the behavior of the
probability density  of equation (\ref{solution-both}), considering
the sign in $-\mu^2$, taking the values $\mu=15$, $Q=0$ and
$\theta=0,0.5, 0.8, 1$, respectively.}\label{c+0}
\end{figure}

\begin{figure}[ht]
\includegraphics[scale=0.6]{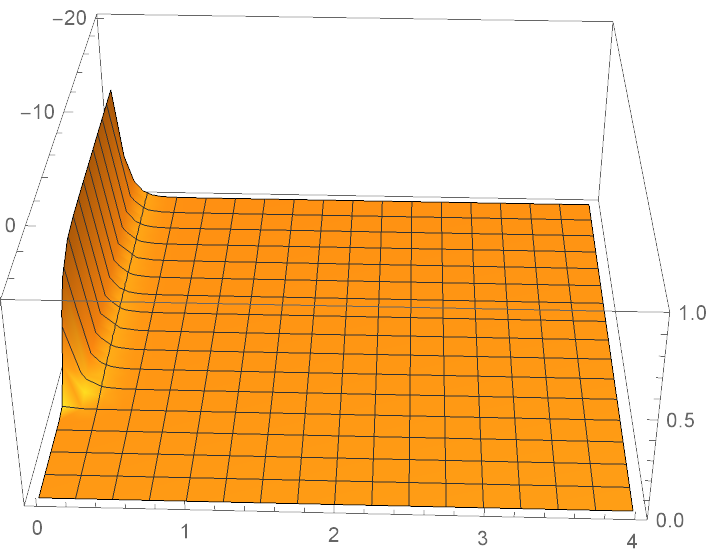}
\includegraphics[scale=0.6]{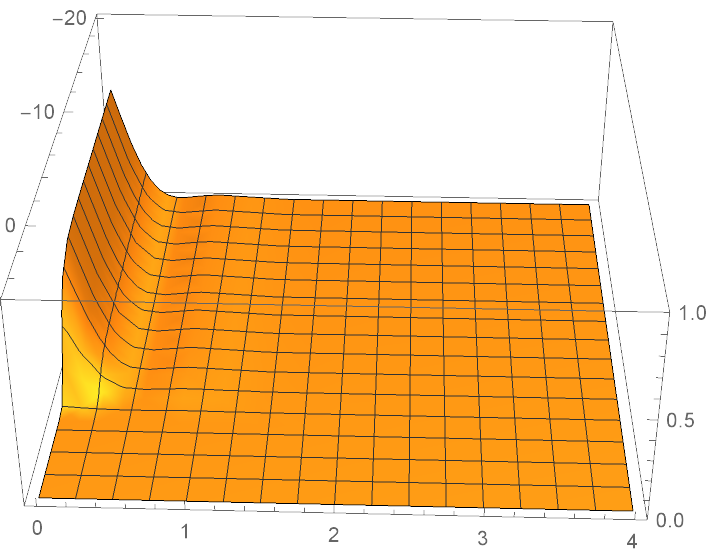}
\includegraphics[scale=0.6]{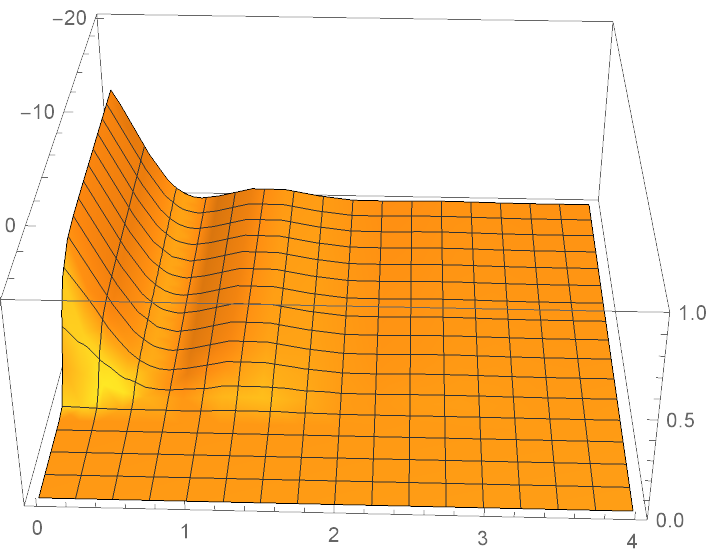}
\includegraphics[scale=0.6]{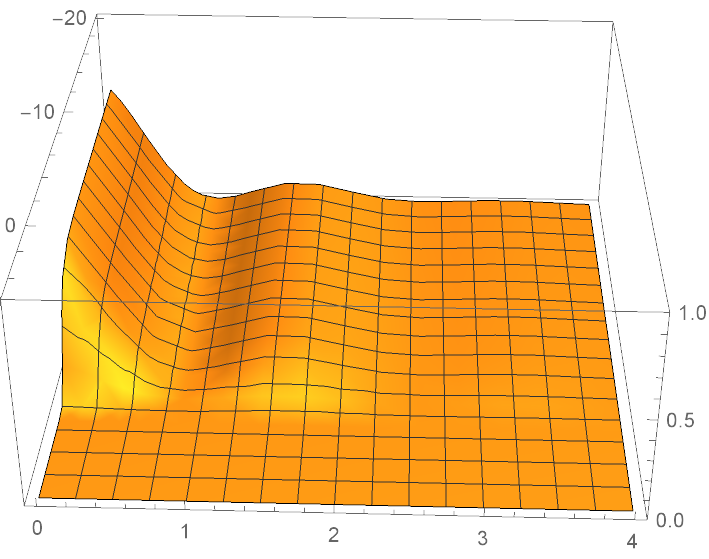}
\caption{In the following plots, we show the behavior of the
probability density of equation (\ref{solution-both}), considering
the sign in $-\mu^2$, taking the values $\mu=15$, $Q=1$ and
$\theta=0,0.5, 0.8, 1$, respectively.}\label{c+1}
\end{figure}

\begin{figure}[ht]
\includegraphics[scale=0.6]{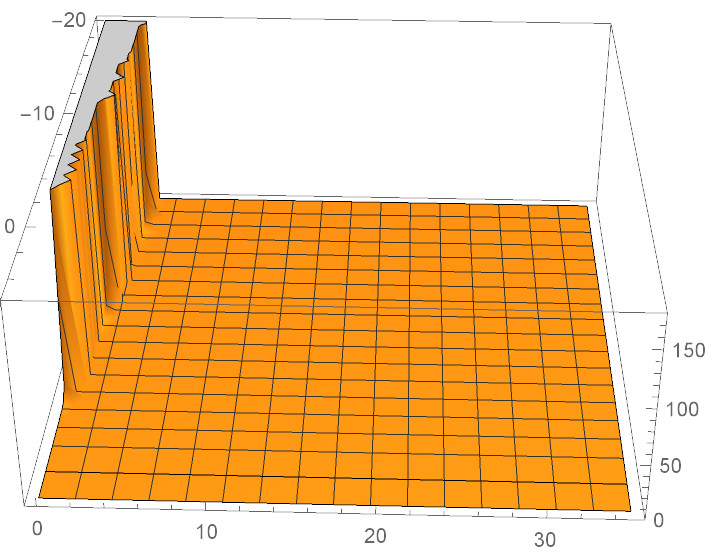}
\includegraphics[scale=0.6]{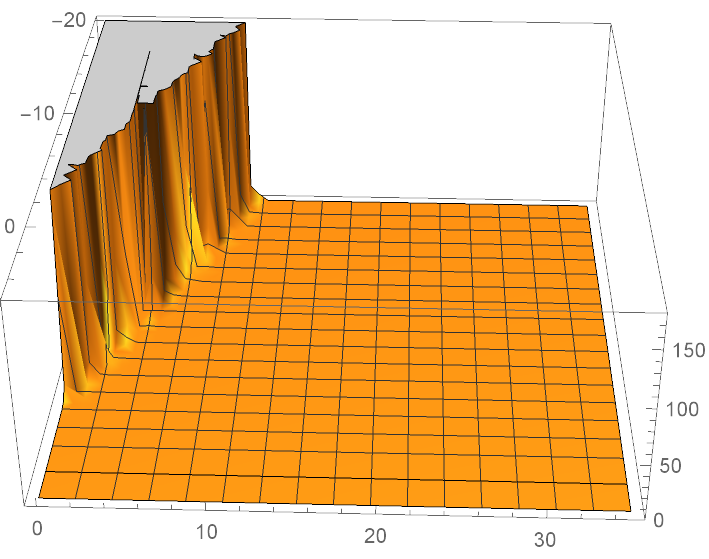}
\includegraphics[scale=0.6]{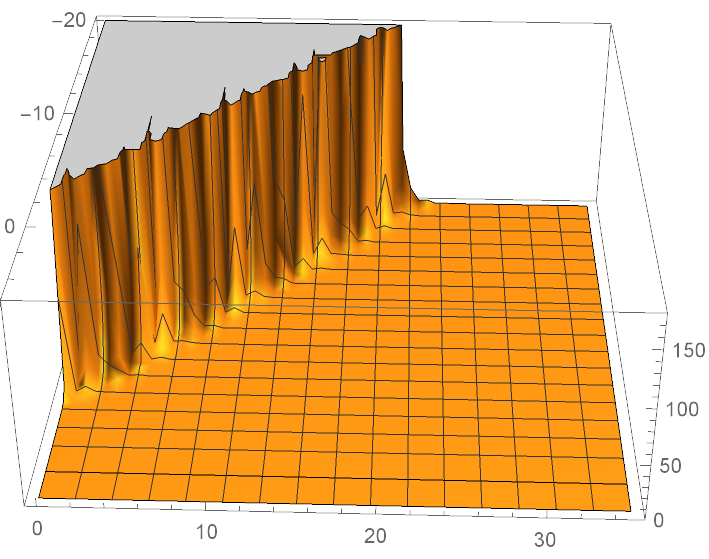}
\includegraphics[scale=0.6]{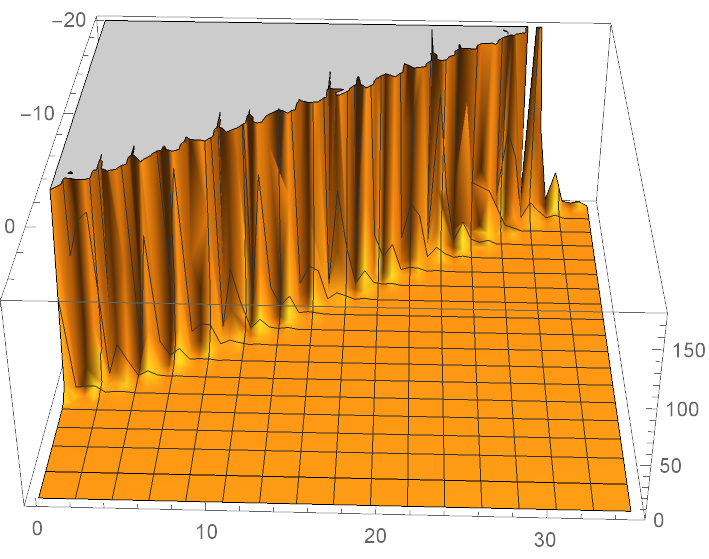}

\caption{In the following plots, we show the behavior of the
probability density of equation (\ref{solution-both}), considering
the sign in $-\mu^2$, taking the values $\mu=15$, $Q=-1$ and
$\theta=0,0.5, 0.8, 1$, respectively.}\label{c+-1}
\end{figure}

For the other scenarios, employing the modified Bessel function, the
behavior is very different, as shown in the figures \ref{c+0},
\ref{c+1} and \ref{c+-1}, when the combination of the parameters
$\mu=15$ and $\theta=0,0.5,0.8, 1$, having decaying behavior in the
direction of evolution of the scale factor like ($\Omega$) and
oscillatory behavior in the direction of the scalar field, or making
the scalar field relevant in quantum evolution and remaining in
classical evolution, as has been found in other alternative models
to Einstein's theory
\cite{omar2017,omar2018,omar2020,abraham2021,barron2021,abraham2022,abraham2023}.

Since, we do not know the initial conditions of the universe in the
dust epoch, we have graphed both probability densities, where it is
observed that the scalar field persists in the evolution of both
densities, remaining as a remnant towards the classical evolution of
the universe, being a cosmic background currently.



The global effect of the non-commutativity  between the field
coordinates of the system  in fractional quantum cosmology scheme
causes the probability density to shift or shrink in the opposite
direction to the scale factor, causing the classical universe to
emerge sooner, which would mean that the current universe should
have more time than is usually mentioned, as in mentioned in the
reference \cite{Jalalzadeh2023}, employing the fractional framework.

On the other hand, if the order of the differential equation
\eqref{ma-ster} is a rational, then solutions have two cases

\subsection{$\omega_x \in [0,1]$, $\beta \in [1,2]$, $n=\lceil \beta \rceil=2$}
Taking into account the Laplace transform in \cite{Machado},
considering that ${\cal L}[_cD^\beta f(t)]=s^\beta F(s)-s^{\beta
-1}f(0)-s^{\beta-2}f^\prime(0)$, ${\cal L}[\frac{d f(t)}{dt}]=s
F(s)- f(0)$, and ${\cal L}[ f(t)]=F(s)$. Then, let the fractional
differential equation
\begin{equation}
\frac{d^{\beta} {\cal C}}{ d \phi^\beta} + A
\frac{d{\cal C}}{d\phi} + B {\cal C}=0, \label{EQ}
\end{equation}
where
\begin{equation}
A = \mp \frac{i\theta}{2}\Big(\frac{\alpha}{2^{\alpha -
1}}\Big)^{\frac{1}{ 2\alpha - 1}}\frac{\mu^2 \beta}{24
\hbar^{2\gamma}}\qquad B = \pm \Big(\frac{\alpha}{2^{\alpha -
1}}\Big)^{\frac{1}{ 2\alpha - 1}}\frac{\mu^2 \beta}{24
\hbar^{2\gamma}}. \label{TE}
\end{equation}
Applying the Laplace transform to all the terms in (\ref{EQ}), we have

\begin{equation}
s^{\beta}C(s) - s^{\beta - 1}C(0) - s^{\beta - 2}C^\prime(0) + A
s\,C(s) - A\,C(0) + B\,C(s) = 0
\end{equation}
Solving with respect to $C(s)$, we get
\begin{equation}
C(s) = \frac{C(0) s^{\beta - 1}}{s^{\beta} + As + B }+
\frac{C^\prime(0) s^{\beta - 2}}{s^{\beta} + As + B }+
\frac{AC(0)}{s^{\beta} + As + B} \label{T}
\end{equation}
for the particular value $\beta=2$, the two last terms can be
consider as one, making that $C^\prime(0)+A C(0)=\kappa =constant$,
and for $\beta=1$, the first and last term can be simplify to $C(0)+
A C(0)=\kappa_1=constant$

From the formula in \cite{Machado} (page 40, equation (3.11) with a
correction), we have
\begin{equation}
{\cal L}^{-1}\Big[\frac{s^{\gamma} }{s^\alpha + as^\beta + b}\Big] =
t^{\alpha - \gamma -1}\sum_{n=0}^\infty \sum_{k=0}^\infty
\frac{(-b)^n (-a)^k \binom{n +1 + k}{k} }{\Gamma[k(\alpha - \beta) +
(n+1)\alpha - \gamma] } t^{k(\alpha - \beta) + n\alpha},
\label{master-equation}
\end{equation}
adapting our parameters to the master equation
(\ref{master-equation}), we have the following three cases
\begin{enumerate}
\item{} first term in (\ref{T}), we use $\gamma=\beta-1$, $\alpha=\beta$, $\beta=1$
\begin{equation}
{\cal L}^{-1}\Big[\frac{s^{\beta -1} }{s^\beta + as + b}\Big] =
 \sum_{n=0}^\infty \sum_{k=0}^\infty
\frac{(-B)^n (-A)^k \binom{n +1 + k}{k} }{\Gamma[k(\beta - 1) +
(n+1)\beta -(\beta-1)] } \phi^{k(\beta - 1) + n\beta}
\end{equation}
\item{}  second term in (\ref{T}), we use $\gamma=\beta-2$, $\alpha=\beta$, $\beta=1$
\begin{equation}
{\cal L}^{-1}\Big[\frac{s^{\beta -2} }{s^\beta + as + b}\Big] = \phi
\sum_{n=0}^\infty \sum_{k=0}^\infty \frac{(-B)^n (-A)^k \binom{n + 1
+k}{k} }{\Gamma[k(\beta - 1) + (n+1)\beta -(\beta-2)] }
\phi^{k(\beta - 1) + n\beta}
\end{equation}
\item{} third term in (\ref{T}), we use $\gamma=0$, $\alpha=\beta$, $\beta=1$
\begin{equation}
{\cal L}^{-1}\Big[\frac{1 }{s^\beta + as + b}\Big] = \phi^{\beta -1}
\sum_{n=0}^\infty \sum_{k=0}^\infty \frac{(-B)^n (-A)^k \binom{n +
1+ k}{k} }{\Gamma[k(\beta - 1) + (n+1)\beta ] } \phi^{k(\beta - 1) +
n\beta}
\end{equation}
\end{enumerate}

Then, the inverse Laplace transform of (\ref{T}), is Then, the
inverse Laplace transform of (\ref{T}), is
\begin{eqnarray}
{\cal C} (\phi) &=& {\cal L}^{-1}\Big[ \frac{C(0) s^{\beta - 1}}{s^{\beta} + As + B } + \frac{C^\prime(0) s^{\beta - 2}}{s^{\beta} + As + B } + Ac_0\frac{1}{s^{2\gamma} + As + B}\Big] = \nonumber\\
&=& C(0)\sum_{n=0}^\infty \sum_{k=0}^\infty \frac{(-B)^n (-A)^k
\binom{n +1+ k}{k} }{\Gamma[k(\beta - 1) + (n+1)\beta -(\beta-1)] }
\phi^{k(\beta - 1) + n\beta} + \nonumber\\
 &+& C^\prime(0) \phi
\sum_{n=0}^\infty \sum_{k=0}^\infty \frac{(-B)^n (-A)^k \binom{n +1+
k}{k} }{\Gamma[k(\beta - 1) + (n+1)\beta -(\beta-2)] } \phi^{k(\beta
- 1) + n\beta} +
 \nonumber\\
&+&  A\,C(0) \phi^{\beta -1} \sum_{n=0}^\infty \sum_{k=0}^\infty
\frac{(-B)^n (-A)^k \binom{n +1+ k}{k} }{\Gamma[k(\beta - 1) +
(n+1)\beta ] } \phi^{k(\beta - 1) + n\beta}
\end{eqnarray}
For the case when $A = \mp i\frac{\theta}{2}q_{\alpha,\gamma}$ and
$B = \pm q_{\alpha,\gamma}$, $q_{\alpha,\gamma} =
\Big(\frac{\alpha}{2^{\alpha - 1}}\Big)^{\frac{1}{ 2\alpha -
1}}\frac{\mu^2 \beta}{24 \hbar^{2\gamma}}$, we see that the complex
solution can be read as

\begin{align}
\mathcal{C} & = a\Biggl\{1 - \frac{\theta^{2}q^{2}}{4\,\Gamma(2\beta - 1)}\phi^{2\beta - 2} \mp \frac{q}{\Gamma(\beta + 1)}\phi^{\beta} \pm \frac{3\theta^{2}q^{3}}{4\,\Gamma(3\beta - 1)}\phi^{3\beta - 2} + \frac{q^{2}}{\Gamma(2\beta + 1)}\phi^{2\beta}\nonumber\\
& - \frac{3\theta^{2}q^{4}}{2\,\Gamma(4\beta - 1)}\phi^{4\beta - 2}
+ \ldots \Biggr\}
+ b\Biggl\{\phi - \frac{3q^{2}\theta^{2}}{4\,\Gamma(2\beta)}\phi^{2\beta - 1} \mp \frac{q}{\Gamma(\beta + 2)}\phi^{\beta + 1} \pm \frac{3\theta^{2}q^{3}}{2\,\Gamma(3\beta)}\phi^{3\beta - 1} \nonumber\\
& + \frac{q^{2}}{\Gamma(2\beta + 2)}\phi^{2\beta + 1} - \frac{5\theta^{2}q^{4}}{2\,\Gamma(4\beta)}\phi^{4\beta - 1} + \ldots \Biggr\} \pm i\Biggl\{a\Biggr[\frac{q\theta}{2\,\Gamma(\beta)}\phi^{\beta - 1} + \frac{3\theta^{3}q^{3}}{8\,\Gamma(3\beta - 2)}\phi^{3\beta - 3}\nonumber\\
& + \frac{\theta q^{2}}{\Gamma(2\beta)}\phi^{2\beta - 1} +
\ldots\Biggr] +b\Biggr[\frac{q\theta}{\Gamma(\beta + 1)}\phi^{\beta}
+ \frac{3q^{2}\theta}{2\,\Gamma (2\beta + 1)}\phi^{2\beta} +
\frac{2q^{3}\theta}{\Gamma(3\beta + 1)}\phi^{3\beta} +
\ldots\Biggl]\Biggl\}, \label{solf1}
\end{align}
where $C(0) = a$ and $C'(0) = b$.

\subsection{$\omega_x \in (0,-1]$, $\beta \in [0,1]$, $n=\lceil \beta \rceil=1$}
For this case, the equation to solve is
\begin{equation}
C(s) = \frac{C(0) s^{\beta-1 }}{s^{\beta} + As + B }+
\frac{AC(0)}{s^{\beta} + As + B}, \label{T0}
\end{equation}
similarly, as in the previous case, we have
\begin{eqnarray}
{\cal C} (\phi) &=& {\cal L}^{-1}\Big[ \frac{C(0) s^{\beta - 1}}{s^{\beta} + As + B }  +
\frac{AC(0)}{s^{\beta} + As + B}\Big] = \nonumber\\
&=& C(0)\sum_{n=0}^\infty \sum_{k=0}^\infty \frac{(-B)^n (-A)^k
\binom{n +1+ k}{k} }{\Gamma[k(\beta - 1) + (n+1)\beta -(\beta-1)] }
\phi^{k(\beta - 1) + n\beta} + \nonumber\\
&+&  A\,C(0) \phi^{\beta -1} \sum_{n=0}^\infty \sum_{k=0}^\infty
\frac{(-B)^n (-A)^k \binom{n +1+ k}{k} }{\Gamma[k(\beta - 1) +
(n+1)\beta ] } \phi^{k(\beta - 1) + n\beta}. \label{cmenos}
\end{eqnarray}
which can be rewritten as

\begin{align}
\mathcal{C}(\phi) & = a\Biggl\{1 - \frac{q^{2}\theta^{2}}{4\,\Gamma(2\beta - 1)}\phi^{2\beta - 2} \mp \frac{q}{\Gamma(\beta + 1)}\phi^{\beta} \pm \frac{3 q^{3}\theta^{2}}{4\,\Gamma(3\beta - 1)}\phi^{3\beta - 2} + \frac{q^{2}}{\Gamma(2\beta + 1)}\phi^{2\beta} \nonumber\\
& - \frac{3q^{4}\theta^{2}}{2\,\Gamma(4\beta - 1)}\phi^{4\beta - 2}
+ \ldots\Biggr\} \pm ia\Biggl\{\frac{q\theta}{2\,\Gamma(\beta
)}\phi^{\beta - 1} +
\frac{3q^{3}\theta^{3}}{8\,\Gamma(3\beta - 2)}\phi^{3\beta - 3} - \frac{q^{2}\theta}{\Gamma(2\beta)}\phi^{2\beta - 1}\nonumber\\
& - \frac{3q^{4}\theta^{3}}{4\,\Gamma(4\beta - 2)}\phi^{4\beta - 3}
+ \frac{3q^{3}\theta}{2\,\Gamma(3\beta)}\phi^{3\beta - 1} -
\frac{5q^{5}\theta^{3}  }{4\,\Gamma(5\beta - 2)}\phi^{5\beta - 3} +
\ldots\Biggr\}, \label{solpf2}
\end{align}

\section{Conclusions}\label{comment}

During the development of research in the non-commutative formalism
within fractional cosmology in $k$-essence theory, the presence of
non-commutativity that usually accompanied the term of the scale
factor, here is disrupted, since essentially Non-commutativity is
more present in the scalar field, modifying the mathematical
structure that usually occurs in works in this direction in other
non-fractional formalisms.

In our non-commutative quantum development, the method of separation
of variables does not appear in a traditional way as the sum of the
operators in their variables, now it is produced as factors, thanks
to this it can be
 separated, in addition, now complex fractional differential equations arise, even in cases with derivatives of integer order,
 which means that these solutions in the scalar field have a real part and an imaginary part.

In previous non-commutative quantum works
\cite{Sabido,Aguero,Socorro}, the term is usually modified with the
scale factor, but in fractional cosmology in essence K, this term
remains unchanged, only the scalar field term undergoes important
modifications, in the sense that the probability density undergoes a
shift back to the direction of the scale factor, causing classical
evolution to arise earlier than in the commutative world. This
effect is due to the non-commutativity between the field coordinates
in this formalism, which is related to some crucial effects due to
 the fact of having a fractional equation, such that the age of the universe is greater, of the order of 13.8196 Gyr. , or more
 \cite{Jalalzadeh2023}.
 These results on fractal K-essence
theory add to the fact that this formalism without considering
ordinary matter is falsified with this approach according to the
classical solutions that are identical using the FRW model
\cite{universe}, but it is found that this is a more general result
mentioned in the reference \cite{sasaki2010}.

Since the prefactor that is usually linked to the ordering of the
factors under a certain gauge does not appear in the standard
quantum Hamiltonian, the important contribution of noncommutativity
appears in the wave function linked to the scale factors, which is
why this term continues to persist. This causes the momentum
associated with the scalar field to produce an additional total
derivative term to the non-commutative fractional equation due to
the Bopp shift in the scale factor term, producing in this case a
significant contribution of the non-commutative parameter in the
wave function, see equation (\ref{ma-ster}).

\acknowledgments{ J.S. was partially supported by PROMEP grants
UGTO-CA-3. Both authors were partially supported by SNI-CONACyT.
J.J. Rosales is supported by the UGTO-CA-20 nonlinear photonics and
Department of Electrical Engineering. L.T.S. is supported by
Secretaria de Investigaci\'on y Posgrado del Instituto Polit\'ecnico
Nacional, grant SIP20211444. This work is part of the collaboration
within the Instituto Avanzado de Cosmolog\'ia.  Many calculations
were done by Symbolic Program REDUCE 3.8.


\bigskip

\begin{thebibliography}{99}
\bibitem{Socorro1} Socorro J., Luis O. Pimentel and Abraham Espinoza Garc\'ia, Advances in High Energy Phys. 805164 (2014),
    {\it Classical Bianchi type I cosmology in K-essence theory}

\bibitem{universe} J. Socorro and J. Juan Rosales, Universe {\bf 9}, 185 (2023),  {\it Quantum fractionary cosmology: K-essence theory},
     [arXiv:2302.07799].

\bibitem{fra-fractionary} Socorro, J.;  Juan Rosales, J.; and Toledo-Sesma, L.  Fractal Fract.  {\bf 7}, 814 (2023),
    {\it Anisotropic fractional cosmology: K-essence    theory},   [arXiv:2308.10381].

\bibitem{Abreu-2006} E.M.C. Abreu, C. Neves, and W. Oliveira, Int. J. Mod. Phys. A {\bf 21}, 5359 (2006)
    {\it Noncommutativity from the symplectic point of view}.

\bibitem{De-andrade} M. A. De Andrade, C. Neves, Journal of Mathematical Physics 59, 012105 (2018).
    {\it Noncommutative Mapping from the symplectic formalism}.

\bibitem{Sabido} W. Guzm\'an, C. Ortiz, M. Sabido, J. Socorro and M. Ag\"uero,
    Int. J. Mod. Phys. D {\bf 16} (10), 1625-1632 (2007),    {\it Noncommutative Bianchi quantum cosmology}.

\bibitem{Aguero} M. Ag\"uero, J. A. Aguilar S., C. Ortiz, M. Sabido and J. Socorro,  Int. J. Theor. Phys.
    {\bf 46} (11) 2928-2934 (2007)    {\it Non Commutative  Bianchi type II Quantum Cosmology}.
    [gr-qc/0703151].


\bibitem{Guzman} W. Guzm\'an, M. Sabido and   J. Socorro, Phys. Rev. D {\bf 76}, 087302 (1-4)
     (2007), {\it Noncommutativity and scalar field cosmology}, [gr-qc/0712.1520].

\bibitem{Ortiz} C. Ortiz, E. Mena, M. Sabido and  J. Socorro, Int. J. Theor. Phys. {\bf 47} (5), 1240-1251 (2008).
    {\it (Non) commutative isotropization in Bianchi I with barotropic perfect fluid
    and $\rm \Lambda$ cosmological}. [gr-qc/0703101].

\bibitem{Socorro} J. Socorro, Luis O. Pimentel, C. Ortiz and M. Aguero, Int. J. Theor. Phys. {\bf 48}, 3567-3585 (2009),
     {\it Scalar field in the  Bianchi I: Non commutative classical and Quantum Cosmology }.
      [arXiv:0910.2449].

\bibitem{Guzman-2011} W. Guzman, M. Sabido and J. Socorro, Phys. Lett. B {\bf 697}, 271-274 (2011),
    {\it On Noncommutative Minisuperspace and the Friedmann equations}, [arXiv:0812.4251].

\bibitem{sabido2018} J.L. L\'opez, M. Sabido and C. Yee-Romero,
Phys. of dark Universe {\bf 19}, 104-108 (2018), {\it Phase space
deformation in phantom cosmology}, [arXiv:1711.01111 (gr-qc)].

\bibitem{sabido2024} J.L. L\'opez Pic\'on, M. Sabido and C. Yee-Romero,
Phys. Lett. B {\bf 849}, 138420 (2024), {\it Phase space deformation
in SUSY cosmology}, [arXiv:2309.0587 (gr-qc)].


\bibitem{1} Espinoza-Garc\'ia, A.; Socorro, J.; Pimentel, L.O. {Quantum Bianchi type IX cosmology in K-essence theory.} \emph{Int. J. Theor. Phys.} \textbf{2014}, \emph{53}, 3066--3077.
https://doi.org/10.1007/s10773-014-2102-0 

\bibitem{roland} de Putter, R.; Linder, E.V. {Kinetic k-essence and Quintessence}. \emph{Astropart. Phys.} \textbf{2007}, \emph{28}, 263.

\bibitem{chiba} Chiba, T.; Dutta, S.; Scherrer, R.J. {Slow-roll k-essence}. \emph{Phys. Rev. D} \textbf{2009}, \emph{80}, 043517.

\bibitem{bose} Bose, N.; Majumdar, A.S. {A k-essence model of inflation, dark matter and dark energy}. \emph{Phys. Rev. D} \textbf{2009}, \emph{79}, 103517.

\bibitem{arroja} Arroja, F.; Sasaki, M. {A note on the equivalence of a barotropic perfect fluid with a k-essence scalar field}. \emph{Phys. Rev. D} \textbf{2010}, \emph{81}, 107301.

\bibitem{tejeiro} Garc\'ia, L.A.; Tejeiro J.M.; Casta\~neda, L. {K-essence scalar field as dynamical dark energy}.
    \emph{arXiv} \textbf{2012}, [arXiv:1210.5259. (gr-qc)]

\bibitem{Szabo2} R.J. Szabo, Phys. Rep. {\bf 378}, 207 (2003), [arXiv:hep-th/0109162].

\bibitem{HH} Hartle, J.B.; Hawking, S.W. {Wave function of the Universe}. \emph{Phys. Rev. D} \textbf{1983}, \emph{28}, 2960--2975.
\bibitem{Rosales1}Rosales, J.J.; G\'omez, J.F.; Gu\'ia, M.; Tkach, V.I., \emph{Fractional electromagnetic waves}, In Proceedings
of the 11th International Conference on Laser and Fiber-Optical
Networks Modeling (LFNM), Kharkov, Ukraine, 5--9  September 2011
https://doi.org/10.1109/LFNM.2011.6144969.

\bibitem{Rosales2} G\'omez Aguilar, J.F.; Rosales, J.J.; Bernal Alvarado, J.J.;
Cordova Fraga, T.; Guzm\'an Cabrera, R. {Fractional mechanics
oscillators}. \emph{Rev. Mex. F\'is.} \textbf{2012}, {\emph 58},
348--352.

\bibitem{polyanin} Polyanin, A.C.; Zaitsev, V.F. \emph{Handbook of Exact
Solutions for Ordinary Differential Equations}, 2nd ed.; Chapman \&
Hall/CRC: Boca Raton, FL, USA, 2003.
\bibitem{ncqc} H. Garcia-Compean, O. Obreg\'on and C. Ram\'{\i}rez,  Phys.
Rev. Lett.  {\bf 88}, 161301 (2002).

\bibitem{pimentel}  L.O. Pimentel and C. Mora,   Gen. Rel. Grav.  {\bf 37},
817 (2005),  [arXiv:gr-qc/0408100].


\bibitem{omar2017} J. Socorro and Omar E. Nu\~{n}ez, Eur. Phys. Journal
Plus {\bf 132}: 168 (2017),
    {\it Scalar potentials with multi-scalar fields from quantum cosmology and supersymetric quantum mechanics},
     [arXiv:1702.00478]

\bibitem{omar2018} J. Socorro  Omar E. N\'u\~nez and Rafael Hern\'andez-Jim\'enez, Advances in Math. Phys.
    Volume (2018), Article ID 3468381, {\it Classical and quantum exact solutions for a FRW multi-scalar field cosmology with an exponential
    potential driven inflation}, [arXiv:1811.11565 in gr-qc].

\bibitem{omar2020} J. Socorro, Omar E. N\'u\~nez, Rafael Hern\'andez-Jim\'enez, Phys. Lett. B {\bf 809}, 135667 (2020),
    {\it Classical and quantum exact solutions for the anisotropic Bianchi type I in multi-scalar field cosmology
    with an exponential potential driven inflation}, [arXiv:1904.00807 in
    gr-qc].

\bibitem{abraham2021} J. Socorro, S. P\'erez-Pay\'an, Rafael Hern\'andez-Jim\'enez, Abraham Espinoza-Garc\'ia and Luis Rey D\'iaz-Barr\'on,
    Class. Quantum Grav. {\bf 38}, 135027 (2021), {\it Classical and quantum exact solutions for a FRW  in chiral like cosmology.} [arXiv:2012.11108,
    (gr-qc)].

\bibitem{barron2021}  Luis Rey D\'iaz-Barr\'on, S. P\'erez-Pay\'an, Abraham Espinoza-Garc\'ia  and J. Socorro, Int. J. Mod. Phys. D {\bf 30}
    (11), 2150080 (2021), {\it Anisotropic chiral cosmology: exact solutions}, [arXiv:2101.05973, (gr-qc)].

\bibitem{abraham2022} J. Socorro, S. P\'erez-Pay\'an, Rafael Hern\'andez-Jim\'enez, Abraham Espinoza-Garc\'ia and Luis Rey D\'iaz-Barr\'on,
    Universe {\bf 8}, 548 (2022), {\it Quintom fields from chiral K-essence cosmology},
    [arXiv:2204.12083].

\bibitem{abraham2023} J. Socorro, S. P\'erez-Pay\'an, Rafael Hern\'andez-Jim\'enez, Abraham Espinoza-Garc\'ia and Luis Rey D\'iaz-Barr\'on,
    General Relativity and Gravitation {\bf 55}: 75,(2023), {\it Quintom fields from chiral anisotropic cosmology },
    [arXiv:2210.01186].
\bibitem{Jalalzadeh2023} Emanuel Wallison de Oliveira Costa, Raheleh Jalalzadeh, Pedro Felix da Silva J\'unior, Seyed Meraj Mousavi Rasouli and Shahram Jalalzadeh,
    Fractal Fract. {\bf 7}, 854 (2023), {\it Estimated Age of the Universe in Fractional Cosmology}.

\bibitem{Machado} Constantin Milici, Gheorghe Draganescu, J. Tenreiro Machado.
{\it Introduction to Fractional Differential Equations}. Springer
Nature Switzerland AG (2019).

\bibitem{sasaki2010} Frederico Arroja and Misao Sasaki, Phys. Rev. D {\bf 81}, 107301 (2010), {\it Note on the equivalence of a
    barotropic perfect fluid with a k-essence scalar field}.

\end{thebibliography}
\end{document}